\documentclass[pra,aps,showpacs,twocolumn,floatfix,superscriptaddress]{revtex4}  
\usepackage{graphicx}
\usepackage[ansinew]{inputenc}
\usepackage{array}
\usepackage{color}
\usepackage{amsmath}
\usepackage{amsxtra}
\usepackage{amstext}
\usepackage{amssymb}
\usepackage{latexsym}
\usepackage{dsfont}
\usepackage{verbatim}
\usepackage{comment}
\usepackage{epstopdf}

\begin{document}

\title
{Divergent nonlinear optical response of three resonator system via Fano resonances}



\author{Mehmet Emre Ta\c{s}g\i n}
\affiliation{Institute of Nuclear Sciences, Hacettepe University, 06800, Ankara, Turkey}

\date{\today}

\begin{abstract}
In a previous study, we have discovered that nonlinear processes can be enhanced several orders of magnitude due to the path interference effects which are introduced by Fano resonances. Emergence of this phenomenon has been demonstrated also in 3-dimensional solutions of Maxwell equations. However, enhancement has been found to be limited by the decay rate of the plasmonic oscillations. In the present work, we demonstrate that such a limitation can be lifted off when the path interference from two quantum emitters to a plasmonic resonator (second harmonic converter) is considered. 
Enhancement factors much larger than 7 orders of magnitude are possible using such an interference scheme. Therefore, a  single hybridized system of 3 particles --manufactured carefully-- can account for almost all of the second harmonic generated radiation emitted from a sample of plasmonic particle clusters decorated with quantum emitters (e.g. a spaser).
\end{abstract}



\maketitle


Metal nanoparticles (MNPs) interact with the optical light much more strongly compared to quantum dots (QDs) or molecules. The localized surface plasmon-polariton (PP) fields concentrate the incident electromagnetic field to small dimensions. Localization results an intensity enhancement as high as $10^5$ \cite{stockman-review,QDinducedtrans} at the hot-spots of MNP interfaces. Such an enhancement in the electromagnetic field also leads nonlinear optical effects to come into play \cite{KauranenNature2012,BouhelierOptExpress2012,DurakNanoLett2007,WalshNanoLett2013,PuPRL2010,WunderlichOptExp2013,YustOptExpress2012,SinghNanotech2013,CentiniOptExpress2011,ZielinskiOptExpress2011,
Gao-AccChemRes-2011}. The emergence of enhanced Raman scattering \cite{Sharma2012}, second harmonic generation (SHG) \cite{,BouhelierOptExpress2012,DurakNanoLett2007,WalshNanoLett2013,PuPRL2010,WunderlichOptExp2013,YustOptExpress2012,SinghNanotech2013,CentiniOptExpress2011,ZielinskiOptExpress2011,
Gao-AccChemRes-2011} and four-wave mixing \cite{GenevetNanoLett2010} can be utilized as optical devices to use in imaging \cite{KneippPRL1997,AngelesNatureNano2010}, as optical switch  \cite{switch} and in generation of quantum entanglement \cite{squeezedSPP,FaselNewJ2006}.

SHG is a process which has a fundamental importance in quantum plasmonics \cite{quantum_plasmonics,plasmonic_goes_quantum,Ozbayplasmonics}. It is responsible for squeezing \cite{squeezing_SHG} which can generate entanglement in nano-dimensions and provides the tool for measurements below the the standard quantum limit (SQL) \cite{measurement_SQL} --which is necessary in quantum plasmonics due to the decrease in signal to noise (SNR) in single-plasmon devices \cite{single_plasmon}. Recent experiments \cite{AltewischerNature2002,LawriePRL2013,squeezedSPP,FaselNewJ2006} revealed that plasmons are able to preserve the quantum entanglement (both in propagation and photon-plasmon conversions \cite{AltewischerNature2002,LawriePRL2013}) much more longer compared to their own lifetime ($\sim 10^{-14}$s). This observation triggered the studies on adoption of plamonic entanglement into quantum information \cite{ChenPRB2011,ChenOptLett2012}. A recent theoretical study \cite{Fanoentanglement} demonstrates also the relevance between the emergence of plasmonic entanglement and Fano resonances \cite{metasginNanoscale2013,QCfano1,QCfano2,QCfano3,QCfano4,QCfano5,QCfano6,QCfano7,QCfano8,QCfano9}.

SHG can also be used in solar cell applications for increasing the absorption efficiency \cite{PV} by up-converting the infrared spectrum to the active frequency region and also for increasing the coherence time of the incident field \cite{BECsqueezed1,BECsqueezed2,metasginNanoscale2013}.

Despite the field enhancement due to surface PP localization \cite{stockman-review,QDinducedtrans}, SHG process still remains weak due to some symmetry requirements \cite{symmetry_selection,CzaplickiPRL2013}. Recently, our group demonstrated \cite{metasginSHG} that Fano resonances can be utilized for the enhancement of the nonlinear response (e.g. SHG) of plasmonic resonators. When a MNP is coupled to a quantum emitter (QE), eg. a QD, the path interference effects can be adopted to cancel the nonresonant frequency terms; thus SHG process can be carried to resonance [see Eq.~(\ref{eq:alph2old})] without modifying the plasmon modes. In Ref.~\cite{metasginSHG}, in order to take retardation effects into account, we performed 3-dimensional simulations which are based on the exact solutions of the Maxwell equations (using MNPBEM toolbox \cite{MNPBEM} in Matlab). We used a Lorentzian dielectric function which has a sharp resonance ($\gamma_{eg}=10^{11}$Hz) to simulate a QD (or a molecule) in MNPBEM toolbox. In these simulations, we obtained parallel results with the simple model of coupled quantum-classical oscillators. 

In a recent experiment conducted by our team \cite{SHG_enhancement_experiment}, we also demonstrate that a 1000 times SHG enhancement is achievable via Fano resonances which enables the observation of a SH signal using a CW laser. Even though our previous studies \cite{metasginSHG,SHG_enhancement_experiment} --where coupling to a single quantum emitter is investigated-- can demonstrate the SHG enhancement upto 2-3 orders of magnitude, this enhancement is limited by the decay rate of the SH plasmon mode [see discussion below Eq.~(\ref{eq:alph2old})]. 

In this paper, we show that such a limitation does not hold anymore, when the path interference of a plasmonic resonator with two (or more) quantum emitters (QEs) is considered. We show that an enhancement factor of 5$\times 10^7$ is easily achievable. The potential for enhancement is expected to be much more above this value. Because, we do not perform a detailed mathematical analysis on the parameter set which maximizes the SHG intensity. Here, we consider coupling of the SH converter (plasmonic non-centrosymmetric dimer) to quantum emitters. However, emergence of conversion enhancement does not necessitate the attachment of high-quality oscillators. Enhancement can emerge even if classical oscillators are used instead of quantum emitters which have sharp resonances [see discussion below Eq.~(\ref{eq:alph2old})]. This is one of the possible reasons for enhancement of SHG from nanoparticle composites and clusters \cite{BouhelierOptExpress2012,DurakNanoLett2007,WalshNanoLett2013,PuPRL2010,WunderlichOptExp2013,YustOptExpress2012,SinghNanotech2013,CentiniOptExpress2011,ZielinskiOptExpress2011,
Gao-AccChemRes-2011}.  

Hence, a  single hybridized system of 3 particles --which is manufactured carefully-- can account for almost all of the second harmonic radiation emitted from a sample of plasmonic particle clusters decorated with quantum emitters (e.g. a spaser). When toy blocks of the self assembled composites of DNA molecules attached to gold nanopaticles \cite{molattached} are considered, such schemes are available within today's technology.

Paper is organized as follows. First, we present the effective Hamiltonian governing the coupled tripartite system. We derive the equations of motion for the oscillations of the plasmon modes. We obtain analytic expression for the steady state value of the plasmon mode($\tilde{\alpha}_2$) where in second harmonic oscillations ($e^{-i2\omega t}$) take place. Second, we discuss the maximization of the SHG amplitude using independent complex variables, e.g. the coupling strengths. We discuss the conversion limitation in the case of a single QE coupled to plasmonic resonator and how this limitation lifted off for the coupling of two QEs to the SH converter (resonator). Last, we summarize our results.

\begin{figure}
\includegraphics[width=3.4in]{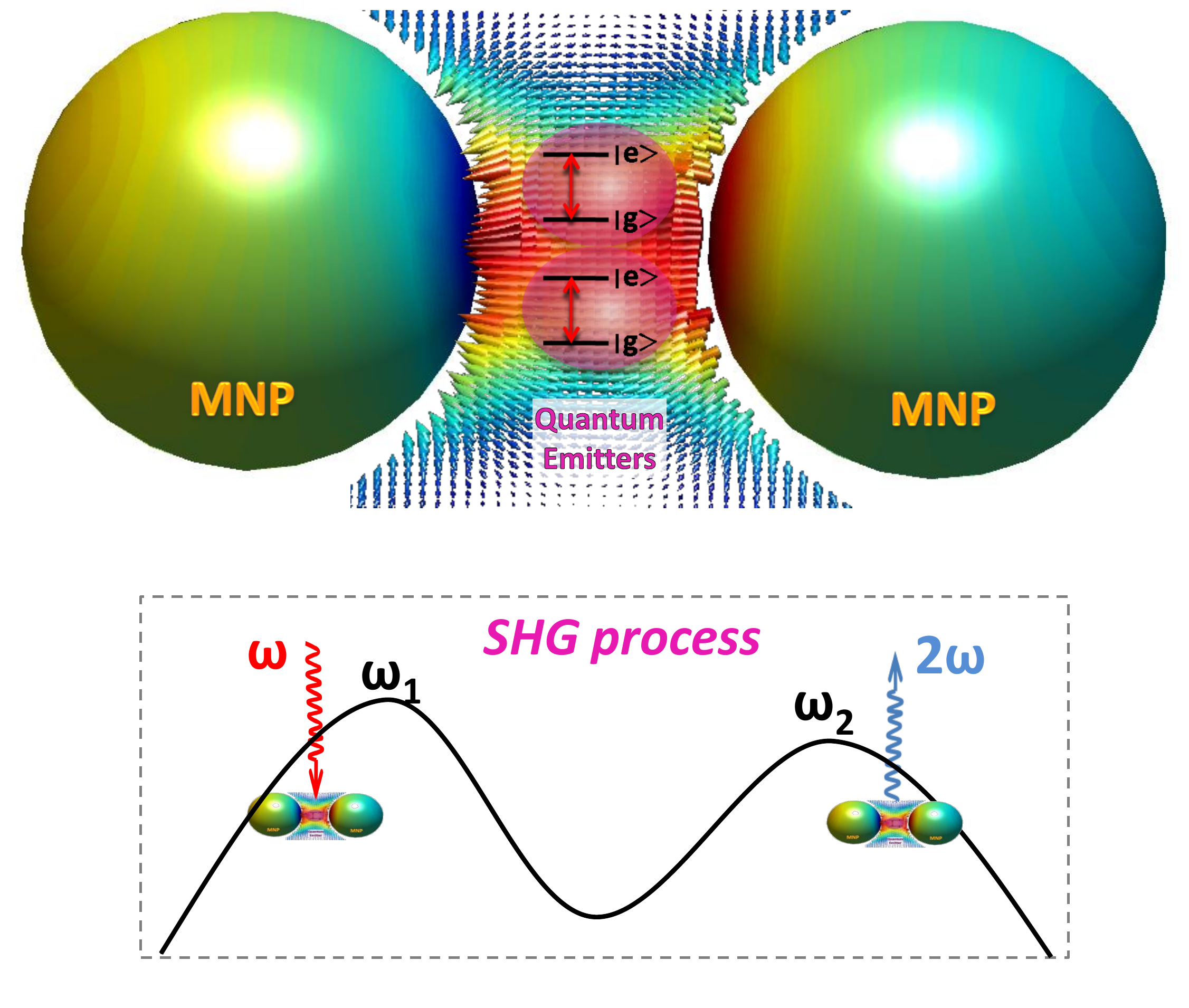}
\caption{(color online) {\it Top:} Two quantum emitters (purple) with small decay rates are placed at the center of a MNP dimer \cite{QCfano1}. The polarization of the plasmon-polariton (PP) modes strongly localizes the incident field to the center (see the field vectors). Field enhancement gives rise nonlinear processes, e.g. second harmonic generation (SHG). The two MNPs are chosen in the same size only for the purposes of demonstration. {\it Bottom:} The incident planewave field ($\epsilon_p e^{-i\omega t}$) drives the $\hat{a}_1$ PP polarization mode (resonance $\omega_1$) of the dimer. The intense localized polarization field of $\hat{a}_1$, oscillating with $\omega$, gives rise to SHG \cite{2plas1phot}. This process induces oscillations ($e^{-i2\omega t}$) in the second PP mode $\hat{a}_2$ whose resonance is $\omega_2$. The quantum oscillators (level spacing $\omega_{eg}^{(1,2)}\simeq 2\omega$) interact with the field of the $\hat{a}_2$ polarization mode. Quantum emitters interact with each other and chosen to have no SH response to frequency $\omega$. The observation of the SH light occurs due to the radiative decay of the $\hat{a}_2$ PP mode \cite{BouhelierPRL2005,2plas1phot,BeversluisPRB2003,PohlScience2005} or fluorescence from the quantum emitters \cite{SHG_enhancement_experiment}.}
\label{fig1}
\end{figure}

\subsection*{Hamiltonian and steady state solutions}

The total Hamiltonian $(\hat{H})$ for the described system can be written as the sum of the energy of the quantum emitters $(\hat{H_0})$, energy of the plasmon-polariton oscillations $(\hat{a_1}\:\text{,}\:\hat{a_2})$ of the MNP dimer $(H_{\rm d})$, the interaction of the quantum emitters (QEs) with the plasmon-polariton modes \cite{metasginNanoscale2013,QCfano1} $(H_{\rm QEs-MNP})$, the interaction of the two quantum emitters with each other ($\hat{H}_{\rm QE-QE}$),
\begin{eqnarray}
\hat{H}_0=\hbar\omega_{\rm eg}^{(1)}| {\rm e_1} \rangle \langle {\rm e_1} | + \hbar\omega_{\rm eg}^{(2)}  | {\rm e_2} \rangle \langle {\rm e_2}| , \label{eq:H0} 
\\
\hat{H}_{\rm d}=\hbar\omega_1\hat{a}_1^\dagger\hat{a}_1+\hbar\omega_2\hat{a}_2^\dagger\hat{a}_2 , \label{eq:Hd}
\\
\hat{H}_{\rm QEs-MNP}=\hbar(f_1\hat{a}_2^\dagger| {\rm g}_1 \rangle \langle {\rm e}_1 |+f_1^*\hat{a}_2| {\rm e}_1 \rangle \langle {\rm g}_1 |) \nonumber \\  
+\hbar(f_2\hat{a}_2^\dagger| {\rm g}_2 \rangle \langle {\rm e}_2 |+f_2^*\hat{a}_2| {\rm e}_2 \rangle \langle {\rm g}_2 |), \label{eq:Hint}
\\
\hat{H}_{\rm QE-QE}=\hbar \big( g|{\rm e}_2 \rangle \langle {\rm g}_2| \otimes  |{\rm g}_1 \rangle \langle {\rm e}_1| \nonumber \\
+g^*|{\rm e}_1 \rangle \langle {\rm g}_1| \otimes  |{\rm g}_2 \rangle \langle {\rm e}_2| \big),
\end{eqnarray}
as well as the energy transferred by the pump source $(\omega)$, $\hat{H}_{\rm p}$ and the second harmonic generation process among the plasmon-polariton fields $(\hat{H}_{\rm sh})$ 
\begin{eqnarray}
\hat{H}_{\rm p}=i\hbar(\hat{a}_1^\dagger\epsilon_p e^{-i\omega t} -\hat{a}_1\epsilon_p^* e^{i\omega t}) , \label{eq:Hp}
\\
\hat{H}_{\rm sh}=\hbar\chi^{(2)}(\hat{a}_2^\dagger\hat{a}_1\hat{a}_1+\hat{a}_1^\dagger\hat{a}_1^\dagger\hat{a}_2) , \label{eq:Hsh}
\end{eqnarray}
respectively \cite{Scullybook,mandelwolf}. In Eq.~(\ref{eq:H0}), $\omega_{eg}^{(1)}$ and $\omega_{eg}^{(2)}$  are the spacing of the excited and ground energy levels for the first and second QE. States $| {\rm e}_1 \rangle$ and $| {\rm e}_2 \rangle$ ($| {\rm g}_1 \rangle$ and $| {\rm g}_2 \rangle$) correspond to the excited (ground) energy levels of the two QEs. $\hat{a}_1$, $\hat{a}_2$ are the plasmon-polariton excitations induced on the MNP dimer and $\hbar\omega_1$, $\hbar\omega_2$ are the corresponding energies for the oscillation modes. $f_1$ and $f_2$ are the coupling matrix element between the field induced by the $\hat{a}_2$ polarization mode of the MNP dimer and the first and second QE, respectively. $g$ is the coupling matrix element between the two QEs. In Eq.~(\ref{eq:Hint}), we neglect the coupling of the QEs to the $\hat{a}_1$ plasmon-polariton mode since we consider that the level spacing of both QEs ($\omega_{eg}^{(1)}\sim 2\omega$, $\omega_{eg}^{(2)}\sim 2\omega$) are off-resonant to the $\omega_1\sim \omega$ mode.

 Eq.~(\ref{eq:Hp}) describes the interaction of the light source (oscillates as $e^{-i\omega t}$) driving the plasmon-polariton mode with smaller resonance frequency $\omega_1$. In Eq.~(\ref{eq:Hsh}), the fields of two excitations in the low-energy plasmon-polariton mode ($\hat{a}_1$) combine to generate the field of a high energy plasmon-polariton mode. Stronger the second harmonic generated plasmon-polariton oscillations, the higher the number of emitted SHG photons $(2\omega)$. Because, $\hat{a}_2$ mode radiatively decay to $2\omega$ photon mode \cite{BouhelierPRL2005,2plas1phot} or can be detected as fluorescence from attached molecules \cite{SHG_enhancement_experiment}, as well. Energy is conserved in the input-output process. The parameter $\chi^{(2)}$, in units of frequency, is proportional to the second harmonic susceptibility of the MNP dimer. The symbol $\otimes$ stands for the direct product of the two Hilbert spaces belonging to the first and second QE.

We note that, one could also treat the SHG process as originating directly from the incident field, e.g. $\hat{H}_{\rm sh}\sim ( \hat{a}_2^{\dagger} \epsilon_p^2e^{-i2\omega t}+c.c. )$. Even though the following results would remain unaffected, physically such a model would be inappropriate. Because, enhanced nonlinear processes emerge due to the electromagnetic field of the localized intense surface plasmon-polariton (polarization) mode \cite{BouhelierPRL2005,2plas1phot}. However, the mode of the incident field ($\omega$) is planewave.


We use the commutation relations (e.g. $i\hbar\dot{\hat{a}}=[\hat{a}\text{,}\hat{H}]$) in driving the equations of motions. We keep operators $\hat{a}_1\text{,}\:\hat{a}_2$ quantum up to a step  in order to avoid incomplete modeling  of the equations of motion. After obtaining the dynamics in the quantum approach, we carry $\hat{a}_1\text{,}\;\hat{a}_2$ to   classical expectation values $\hat{a}_1\rightarrow\alpha_1\text{,}\;\hat{a}_2\rightarrow\alpha_2$. We introduce the decay rates for plasmon-polariton fields $\alpha_1$, $\alpha_2$. Quantum objects are treated within the density matrix approach. The equations take the form 
\begin{subequations}
\begin{align}
\dot{\alpha}_1=(-i\omega_1-\gamma_1)\alpha_1-i2\chi^{(2)}\alpha_1^*\alpha_2 + \epsilon_p e^{-i\omega t} , \label{eq:timea}
\\
\dot{\alpha}_2=(-i\omega_2-\gamma_2)\alpha_2-i\chi^{(2)}\alpha_1^2-i f_1\rho_{ge}^{(1)}-i f_2\rho_{ge}^{(2)} , \label{eq:timeb}
\\
\dot{\rho}_{ge}^{(1)}=(-i\omega_{eg}^{(1)}-\gamma_{eg}^{(1)})\rho_{ge}^{(1)}+i f_1^*\alpha_2(\rho_{ee}^{(1)}-\rho_{gg}^{(1)}) \nonumber 
\\ +ig^*(\rho_{ee}^{(1)}-\rho_{gg}^{(1)}) \rho_{ge}^{(2)} , \label{eq:timec}
\\
\dot{\rho}_{ge}^{(2)}=(-i\omega_{eg}^{(2)}-\gamma_{eg}^{(2)})\rho_{ge}^{(2)}+i f_2^*\alpha_2(\rho_{ee}^{(2)}-\rho_{gg}^{(2)}) \nonumber 
\\ +ig^*(\rho_{ee}^{(2)}-\rho_{gg}^{(2)}) \rho_{ge}^{(1)} , \label{eq:timed}
\\
\dot{\rho}_{ee}^{(1)}=-\gamma_{ee}^{(1)}\rho_{ee}^{(1)}+i \left( f_1\alpha_2^*\rho_{ge}^{(1)}- f_1^*\alpha_2{\rho_{ge}^{(1)}}^*\right) \nonumber \\
+i\left( g{\rho_{ge}^{(2)}}^*\rho_{ge}^{(1)} - g{\rho_{ge}^{(2)}} {\rho_{ge}^{(1)}}^* \right)
 , \label{eq:timee}
 \\
 \dot{\rho}_{ee}^{(2)}=-\gamma_{ee}^{(2)}\rho_{ee}^{(2)}+i \left( f_2\alpha_2^*\rho_{ge}^{(2)}- f_2^*\alpha_2{\rho_{ge}^{(2)}}^*\right) \nonumber \\
+i\left( g{\rho_{ge}^{(1)}}^*\rho_{ge}^{(2)} - g{\rho_{ge}^{(1)}} {\rho_{ge}^{(2)}}^* \right)
 , \label{eq:timef}
\end{align}
\end{subequations}
where $\gamma_1$, $\gamma_2$ are the damping rates of the MNP dimer modes $\alpha_1$, $\alpha_2$. $\gamma_{ee}^{(1)}$, $\gamma_{ee}^{(2)}$ and $\gamma_{eg}^{(1)}=\gamma_{ee}^{(1)}/2$, $\gamma_{eg}^{(2)}=\gamma_{ee}^{(2)}/2$ are the diagonal and off--diagonal decay rates of the first and second quantum emitter, respectively. To make a comparison, $\gamma_1$,$\gamma_2\sim 10^{14}$Hz for MNPs \cite{QDinducedtrans} while $\gamma_{ee}\sim 10^{12}$Hz for molecules \cite{spaser} and $\gamma_{ee}\sim 10^9$ Hz for quantum dots \cite{QCfano3}. The two constraints regarding the conservation of probability $\rho_{ee}^{(1)}+\rho_{gg}^{(1)}=1$ and $\rho_{ee}^{(2)}+\rho_{gg}^{(2)}=1$ accompany Eqs.~(\ref{eq:timea}-\ref{eq:timef}).



In our simulation, in determining the enhancement factor, we time-evolve Eqs.~(\ref{eq:timea}-\ref{eq:timef}) numerically to obtain the long time behaviors of $\rho_{eg}^{(1)}$, $\rho_{eg}^{(2)}$, $\rho_{ee}^{(1)}$,$\rho_{ee}^{(2)}$, $\alpha_1$, and $\alpha_2$. We determine the values to where they converge when the drive is on for long enough times. We perform this evolution for different parameter sets ($f_1$,$f_2$,$g$,$\omega_{eg}^{(1)}$,$\omega_{eg}^{(2)}$)  with the initial conditions $\rho_{ee}^{(1)}(t=0)=\rho_{ee}^{(2)}(t=0)=0$, $\rho_{eg}^{(1)}(0)=\rho_{eg}^{(2)}(0)=0$, $\alpha_1(0)=0$, $\alpha_2(0)=0$. 

Beside the time-evolution simulations, one may gain understanding about the linear behavior of Eqs.~(\ref{eq:timea}-\ref{eq:timef}) by seeking solutions of the form
\begin{eqnarray}
&\alpha_1(t)=\tilde{\alpha}_1 e^{-i\omega t} \quad , \quad \alpha_2(t)=\tilde{\alpha}_2 e^{-i2\omega t} ,& \nonumber
\\
&\rho_{eg}^{(1)}(t)=\tilde{\rho}_{eg}^{(1)} e^{-i2\omega t} \quad , \quad \rho_{eg}^{(2)}(t)=\tilde{\rho}_{eg}^{(2)} e^{-i2\omega t} ,& \nonumber
\\
&\rho_{ee}^{(1)}(t)=\tilde{\rho}_{ee}^{(1)} \quad , \quad \rho_{ee}^{(2)}(t)=\tilde{\rho}_{ee}^{(2)} ,& 
\label{eq:slowvary}
\end{eqnarray}   
for the steady states of the oscillations. In our numerical simulations governing the time-evolution of Eqs.~(\ref{eq:timea}-\ref{eq:timef}), we check that the solutions indeed converge to the form of Eq.~(\ref{eq:slowvary}) for long-time behavior. 

Inserting Eq.~(\ref{eq:slowvary}) into Eqs.~(\ref{eq:timea}-\ref{eq:timef}), one obtains the equations for the steady state
\begin{subequations}
\begin{align}
[i(\omega_1-\omega)+\gamma_1]\alpha_1+i2\chi^{(2)}\alpha_1^*\alpha_2=\epsilon_p , \label{eq:steadya}
\\
[i(\omega_2-2\omega)+\gamma_2]\alpha_2+i\chi^{(2)}\alpha_1^2=-i f_1{\tilde{\rho}}_{ge}^{(1)} -i f_2\tilde{\rho}_{ge}^{(2)} , \label{eq:steadyb}
\\
[i(\omega_{eg}^{(1)}-2\omega)+\gamma_{eg}^{(1)}]\tilde{\rho}_{ge}^{(1)}=i f_1^*\alpha_2 y_1 +ig^* y_1 \tilde{\rho}_{ge}^{(2)}  , \label{eq:steadyc}
\\
[i(\omega_{eg}^{(2)}-2\omega)+\gamma_{eg}^{(2)}]\tilde{\rho}_{ge}^{(2)}=i f_2^*\alpha_2 y_2 +ig^* y_2 \tilde{\rho}_{ge}^{(1)}  , \label{eq:steadyd}
\\
\gamma_{ee}^{(1)} {\tilde{\rho}}_{ee}^{(1)}=i \left(f_1\alpha_2^*{\tilde{\rho}}_{ge}^{(1)}-f_1^*\alpha_2{{\tilde{\rho}}_{ge}^{* \; (1)}} \right)  \nonumber \\
+i\left( g{{\tilde{\rho}}_{ge}^{* \;(2)}} {\tilde{\rho}}_{ge}^{(1)} - g^*{{\tilde{\rho}}_{ge}^{(2)}} {{\tilde{\rho}}_{ge}^{* \;(1)}} \right), \label{eq:steadye}
\\
\gamma_{ee}^{(2)}{\tilde{\rho}}_{ee}^{(2)}=i \left(f_2\alpha_2^*{\tilde{\rho}}_{ge}^{(2)}-f_1^*\alpha_2{{\tilde{\rho}}_{ge}^{* \;(2)}} \right)  \nonumber \\
+i\left( g{{\tilde{\rho}}_{ge}^{* \;(1)}} {\tilde{\rho}}_{ge}^{(2)} - g^*{{\tilde{\rho}}_{ge}^{(1)}} {{\tilde{\rho}}_{ge}^{* \;(2)}} \right), \label{eq:steadyf}
\end{align}
\end{subequations}
where $\tilde{\alpha}_1$, $\tilde{\alpha}_2$, ${\tilde{\rho}}_{ge}^{(1)}$, ${\tilde{\rho}}_{ge}^{(2)}$, ${\tilde{\rho}}_{ee}^{(1)}$, and ${\tilde{\rho}}_{ee}^{(2)}$ are constants independent of the time. $y_i={\tilde{\rho}}_{ee}^{(i)}-{\tilde{\rho}}_{gg}^{(i)}$ are population inversion ($i=1,2$) for QEs.

Using Eqs. (\ref{eq:steadyc}) and (\ref{eq:steadyd}) in Eq.~(\ref{eq:steadyb}), one can obtain the steady state value of $\hat{a}_2$ plasmon--polariton mode field as
\begin{widetext}
\begin{equation}
\tilde{\alpha}_2= \frac{ i\chi^{(2)}\left(\beta_1\beta_2+y_1y_2{g^*}^2 \right)}{ \left( y_1|f_1|^2\beta_2+y_2|f_2|^2\beta_1 \right) +iy_1y_2g^*(f_1f_2^*-f_2f_1^*) 
-\xi_2\left( \beta_1\beta_2+y_1y_2{g^*}^2 \right) }  \tilde{\alpha}_1^2,
\label{alph2}
\end{equation}
\end{widetext}
where short hand notations are $\xi_1=\left[ i(\omega_1-\omega)+\gamma_1 \right]$, $\xi_2=\left[ i(\omega_2-2\omega)+\gamma_2 \right]$, $\beta_1=\left[ i(\omega_{eg}^{(1)}-2\omega) +\gamma_{eg}^{(1)} \right]$ and $\beta_2=\left[ i(\omega_{eg}^{(2)}-2\omega) +\gamma_{eg}^{(2)} \right]$.

\subsection*{Super enhancement of SHG} \label{sec:suppression}

{\it i) Single quantum emitter case:}

In the case of a single QE coupled to the SH converter resonator, $f_2=g=0$ and $f_1=f_c$, one obtains the SHG plasmon-polariton amplitude \cite{metasginSHG}
\begin{equation}
\tilde{\alpha}_2^{\rm (single\; QE)}=\frac{i\chi^{(2)}}{\frac{|f_c|^2 y}{i(\omega_{eg}-2\omega)+\gamma_{eg}}-[i(\omega_2-2\omega)+\gamma_2]}\;\tilde{\alpha}_1^2 .
\label{eq:alph2old}
\end{equation}
In the denominator of Eq.~(\ref{eq:alph2old}), the nonresonant term ($\omega_2-2\omega$) can be canceled by the imaginary part of $F=|f_c|^2y/[i(\omega_{eg}-2\omega)+\gamma_{eg}]$ if $\omega_{eg}$ and $f_c$ tuned carefully. However, the same method cannot be performed over the $\gamma_2$ term; since real part of $F$ has the same sign with $\gamma_2$ (due to the negative values $y$ assigns). Therefore, enhancement of the SHG using a single quantum emitter (or a single classical emitter) is limited with the resonance value of SH conversion (that is for $\omega_2=2\omega$) which is determined by $\gamma_2$ in Eq.~(\ref{eq:alph2old}).

On the other hand, enhancement of SHG can occur also using a classical oscillator instead of a quantum emitter. In the denominators of both Eq.~(\ref{alph2}) and Eq.~(\ref{eq:alph2old}), cancellation of the nonresonant term $(\omega_2-2\omega)$ does not require small $\gamma_{eg}^{(1,2)}$. Once the coupling between the oscillators ($f_c$) is significantly large, enhancement can take place.

{\it ii) Two quantum emitters case:}

In difference to single QE case, the denominator of Eq.~(\ref{alph2}) can (in principle) be arranged down to very low values in order to enhance $\tilde{\alpha}_2$ to much higher values. In this case, denominator has 3 complex ($f_1$, $f_2$, $g$) and 2 real ($\omega_{eg}^{(1)}$, $\omega_{eg}^{(2)}$) parameters which can be tuned independently.

Obtaining the optimum parameter set ($f_1$,$f_2$,$g$,$\omega_{eg}^{(1)}$,$\omega_{eg}^{(2)}$), for which the maximum value of $|\tilde{\alpha}_2|$ emerges, involves some variational calculus. Numerically, what one needs to perform is the following. A maximization algorithm for 5 variables must be accompanied with a function which solves $\tilde{\alpha}_2$ at each step from the nonlinear Eqs.~(\ref{eq:steadya}-\ref{eq:steadyf}). We fail in using this method. Because, our subroutine fails to produce the correct solutions for Eqs.~(\ref{eq:steadya}-\ref{eq:steadyf}), in which subroutine suggests exactly the trial (starting) values (that we enter) as solutions. 

Alternatively, we maximize the coefficient of $\tilde{\alpha}_1^2$ in Eq.~(\ref{alph2}) by setting $y_1=y_2=-1$ and constraining $f_1$ and $f_2$ to real and equal values. Parameter set obtained in this approach is naturally far away from the optimum one. Because, at resonances $y_1$ and $y_2$ differentiates from -1 substantially, since emitters are excited. Nevertheless, such a very crude treatment results a $5\times 10^{7}$ times SHG enhancement. Therefore, the original optimum set of parameters is expected to yield astronomically large enhancement factors.

We obtain the $5\times 10^{7}$ times enhancement by comparing the steady state values of $|\tilde{\alpha}_2|^2$ --that is the number of SH plasmons in $\hat{a}_2$ mode-- calculated from the time evolution of Eqs.~(\ref{eq:timea})-(\ref{eq:timef}). The parameter set must be adjusted to $f_1=f_2=-0.0994$,  $g=0.0066-i0.0360$, $\omega_{eg}^{(1)}=2.111$ and $\omega_{eg}^{(2)}=2.571$; for the physical system $\omega_1=1$, $\omega_1=2.1$, $\gamma_1=\gamma_2=0.01$, $\gamma_{ee}^{(1)}=\gamma_{ee}^{(2)}=10^{-5}$, where all frequencies are scaled with the excitation frequency $\omega$. The steady state values of the inversions $y_1=\rho_{ee}^{(1)}-\rho_{gg}^{(1)}=-0.998$ and $y_2=\rho_{ee}^{(2)}-\rho_{gg}^{(2)}=-0.764$ differentiate significantly from -1. The dipole excitation for the QEs approaches $|\rho_{eg}^{(1)}|=0.033$ and $|\rho_{eg}^{(1)}|=0.322$ at the steady state.

The following important issue must be commented on, since we announce enhancement factors as high as 8 orders of magnitude. In Eq.~(\ref{eq:Hint}), we neglect the coupling of the low-energy plasmon mode ($\hat{a}_1$) to the QEs due to the presence of far off-resonance. Regarding such fine tunings in the denominator of Eq.~(\ref{alph2}), in some cases, this negligence can be important. However, when this interaction is included in the Hamiltonian~(\ref{alph2}); i) one cannot obtain an intuitive analytic expression as in Eq.~(\ref{alph2}) and ii) time behaviors in Eq.~(\ref{eq:slowvary}) are not valid anymore. Hopefully, comparison of the enhancement factors obtained in Ref.~\cite{metasginSHG} ($\hat{a}_1$ coupling neglected) and Ref.~\cite{SHG_enhancement_experiment} ($\hat{a}_1$ coupling included) shows that: inclusion of interaction with $\hat{a}_1$ mode results larger enhancement values (compare 40 and 1000). This is because, presence of $\hat{a}_1$ coupling introduces additional interference paths which provide wider control over the system.

\subsection*{Conclusions} 

We investigate the nonlinear response of a coupled system of two quantum emitters and a plasmonic resonator (SH converter). Plasmonic resonator possesses second harmonic response and the quantum emitters do not have two-photon absorption mechanism. We demonstrate that second harmonic generation can be enhanced over 7 orders of magnitude. Such an enhancement can be achieved by carefully choosing the strengths of inter-particle interactions and the energy level spacing for quantum emitters.


When a plasmonic resonator (SH converter) is coupled to a single emitter, path interference due to Fano resonances can carry the nonlinear response utmost to its resonance value (i.e. at $\omega_2=2\omega$). However, the path interference from 3 particles is shown to be not limited with such restrictions. Phenomenon can be extended also to other nonlinear conversion processes as discussed in Ref.~\cite{metasginSHG}.

Although we consider the interference of a plasmonic resonator with two quantum emitters here, emergence of the phenomenon does not necessitate high-quality oscillators with sharp resonances. Thus, the phenomenon introduced in this paper is possibly the mechanism responsible for enhanced SHG from nanoparticle composites and clusters \cite{BouhelierOptExpress2012,DurakNanoLett2007,WalshNanoLett2013,PuPRL2010,WunderlichOptExp2013,YustOptExpress2012,SinghNanotech2013,CentiniOptExpress2011,ZielinskiOptExpress2011,
Gao-AccChemRes-2011}.


\begin{acknowledgements}
I acknowledge  support  from T\"{U}B\.{I}TAK-KAR\.{I}YER  Grant No.  112T927. This work was undertaken while I was a guest researcher in Bilkent University with the support provided by  O\u{g}uz G\"{u}lseren.
\end{acknowledgements}


%
%
%
\end{document}